\documentclass[twoside]{sfm}

\input{sf.def}

\begin{document}

\def\llm{{\sc LLmodels}}
\def\atl{{\sc ATLAS9}}
\def\aatl{{\sc ATLAS12}}
\def\starsp{{\sc STARSP}}
\def\aur{$\Theta$~Aur}
\def\logg{\log g}
\def\tauros{\tau_{\rm Ross}}
\def\kms{km\,s$^{-1}$}
\def\bz{$\langle B_{\rm z} \rangle$}
\def\degr{^\circ}
\def\aaps{A\&AS}
\def\aap{A\&A}
\def\apjs{ApJS}
\def\apj{ApJ}
\def\rmxaa{Rev. Mexicana Astron. Astrofis.}
\def\mnras{MNRAS}
\def\actaa{Acta Astron.}
\newcommand{\Tef}{T$_{\rm eff}$~}
\newcommand{\Vt}{$V_t$}
\newcommand{\CC}{$^{12}$C/$^{13}$C~}
\newcommand{\CDC}{$^{12}$C/$^{13}$C~}

\pagebreak

\thispagestyle{titlehead}

\setcounter{section}{0}
\setcounter{figure}{0}
\setcounter{table}{0}

\renewcommand{\vec}[1]{\mbox{\boldmath$#1$}}
\markboth{R. Arlt}{Generation and evolution of stable stellar magnetic fields}

\titl{Generation and evolution of stable stellar magnetic fields
in young A-type stars}{Arlt R.$^1$}
{$^1$Leibniz-Institut f\"ur Astrophysik Potsdam (AIP), An der Sternwarte 16, D-14482 Potsdam, Germany, email: {\tt rarlt@aip.de}
}

\abstre{
While the presence of magnetic fields on low-mass stars is
attributed to a dynamo process essentially driven by convective
motions, the existence of magnetic fields on intermediate-mass 
stars has very likely other reasons. Presuming that the
fields we see are nearly constant in time, the paper focuses on
the generation of stable magnetic configurations at the early stages
of stellar evolution. The convective processing of an initial magnetic 
field during the pre-main-sequence phase is studied in a very simple 
model star. Azimuthal magnetic fields are found to be typical remnants
in the upcoming radiative envelope after the convection has receded.
}
\baselineskip 12pt

\section{Introduction}
The observational evidence of magnetic fields on intermediate-mass 
stars is reviewed elsewhere in this issue. The vast majority of them 
appear to be invariable. Nevertheless, they must change at least 
during relatively short intervals on the evolutionary time-scale 
as the star does change its state at some points during  lifetime.

The following paper will deal with the evolution of magnetic
fields in radiative envelopes into stable configurations resembling 
the observed geometries. The second part will illustrate first
simulations of the processing of an initial magnetic field in the
convective phase of an intermediate-mass star and the implications
for the remnants of that phase.

\section{Stable magnetic configurations}
Intuitively, a good option for very slowly evolving magnetic 
fields are force-free fields for which currents are parallel 
to the magnetic field lines everywhere in the star (and possibly 
also outside the star). In such a situation zero Lorentz forces 
do not cause any flows driven by magnetic fields. Since finite 
conductivity of the stellar plasma leads to some dissipation of 
the currents, the fields will change on the Ohmic time-scale 
which is extremely long given the microscopic plasma magnetic 
diffusivity  of the order of $10^2$--$10^4$ cm$^2$/s (Spitzer 
1962). A structure of a typical spot size of $10^10$~cm faces a 
microscopic diffusion time of about 1~Gyr. 

One solution for stable, force-free fields was found by Chandrasekhar 
\& Kendall (1957) which are stable even at the presence of diffusion. 
The constraint though is that the stability only holds true if the 
sphere is contained in a perfectly conducting medium with zero exterior 
fields. A vacuum condition on the sphere's surface destroys the 
stability of that specific solution and the Chandrasekhar--Kendall 
functions are subject to a diffusive instability.

\section{Evolution into quasi-stable configurations }
If force-free fields are the topologies we observe, magnetic fields 
must have evolved into those stable configurations at some stage 
of stellar evolution, most likely very early on. It is 
very difficult to draw a complete theoretical picture starting 
from the first stages of star formation, which is connected inevitably 
with magnetic fields, including accretion and early convection, to the 
final settling of the intermediate-mass star.

A few authors have addressed the evolution of initial conditions 
into quasi-stable magnetic configurations in numerical simulations. 
Braithwaite \& Nordlund (2006) started with the generation of a 
helical, quasi-stable field with roughly equal poloidal and toroidal 
field strengths. This is compatible with the findings of enhanced 
stability of such combinations by Wright (1973) which were later 
expanded by Braithwaite (2009). Later in the simulations by Braithwaite
\& Nordlund, diffusion gradually opens poloidal field lines to the 
exterior, thereby reducing the space for toroidal field inside the 
star and giving rise to instability. The study was extended by 
Braithwaite (2008) as to generate predominantly non-axisymmetric 
magnetic fields, starting from a random field distribution mimicking 
the remnants from an early convective evolutionary stage. The evolution 
into the stable topology takes place on a few Alfv\'{e}n time-scales, 
while the final configuration changes very slowly on an Ohmic time-scale.

An alternative route has been drawn by Arlt \& R\"udiger (2011a,b) who use 
an initially differentially rotating star to generate strong toroidal 
magnetic fields in the radiative zone. Such fields are prone to the Tayler 
instability predominantly delivering non-axisymmetric configurations. The 
initial differential rotation is presumed to be due to rotational braking of 
the star (theoretically by St\c{e}pie\'n 2000; observationally by Alecian
et al.\ 2013c). The braking may open a route to the discrimination
between normal A stars and Ap stars since only the ones with enormous
braking (a) are slow like Ap stars and (b) have strong enough toroidal
fields to go through the Tayler instability and show non-axisymmetric fields.
(Since strong braking probably requires strong fields, the concept may
be circular reasoning though! -- See Sect.~\ref{convection} for an alternative.) 
The magnetic-field and velocity fluctuations generated by the instability 
reduce the differential rotation in the star and the toroidal fields are 
no longer sustained. The non-axisymmetric perturbations which have grown 
up to this point are then subject only to Ohmic diffusion. The scenario 
has been confirmed by fully compressible simulations by Szklarski \& Arlt
(2013) showing that a given braking of the star may lead to a non-axisymmetric
instability of internal toroidal fields.

The observations by Alecian et al. (2013a) of a rapid field change in a 
Herbig Ae star may be interpreted such that the aligned dipole observed until
2009 is a remnant from the convective dynamo of the pre-main-sequence life
(a shell is enough to provide this), while in 2010/11, the instability set in 
and created non-axisymmetric fields which have been observable since. They
have become invariable, since no dynamo any longer sustains a field that 
can become unstable. The instability remnants are non-axisymmetric and decay
on the very long Ohmic time-scale.

Both types of simulations -- the ones by Braithwaite et al.\ and the ones by
Arlt et al.\ --  show a relatively quick change of magnetic-field 
topology on the Alfv\'{e}n time-scale and end with very, very slowly 
changing non-axisymmetric fields as to represent Ap star magnetic fields.
The actual scenarios are not too different; it is mostly the creations of
initial conditions that make the simulations differ.

\section{Dynamos in Ap stars}
Theoretically, stationary magnetic fields can also be maintained by a 
dynamo process, i.e. by the sustained induction through flows. If 
certain differential rotation and meridional circulation are continuously 
generated by radiation in the radiative envelope, we may assume the flow 
as given and study possible dynamo action. A simple axisymmetric flow 
is in principle able to drive a dynamo (Dudley \& James 1989). 

The flows, however, need to be specially designed: the circulation used by 
Dudley \& James (1989), for example, was not consistent with the differential 
rotation they imposed. A consistent circulation would have the opposite 
orientation and does not drive a dynamo. Also, as soon as a convective 
(diffusive) core not penetrated by the flow is introduced, dynamo action 
becomes much more difficult  to be excited (Arlt 2008). While these simple 
flows deliver simple field topologies they would probably not explain the 
diversity of surface field patterns found on Ap stars, even if dynamo action 
is feasible. The k.o. criterion is the growth time which, for such `laminar 
dynamos', is on the time-scale of Ohmic dissipation as compared to turbulent 
dynamos growing on much shorter time-scales.

A turbulent dynamo may indeed be at play in the convective core of the star 
and has been studied in a number of simulations. While a large-scale poloidal 
field can be generated to thread the radiative zone, the field strength are 
below kGauss-level already at 0.3 stellar radii (the top of what was possible 
to be computed in a global domain, cf. Brun et al. 2005). An additional
imposed field mimicking the effect of a fossil field led to a more efficient
dynamo action and super-equipartition fields (as compared to the convective
motions) in the convective core, reaching 300~kG strength (Featherstone 
et al. 2009). Further studies will be needed to show whether theses fields
can deliver the spots on the surfaces of Ap stars. Given the stability of the
observed spots, the time-scales on which the simulated flux-rise will change 
spots can be a decisive criterion for the concept of a core dynamo providing 
most of the magnetic flux.

\begin{figure}
\begin{center}
\hbox{ \includegraphics[width=8.8cm]{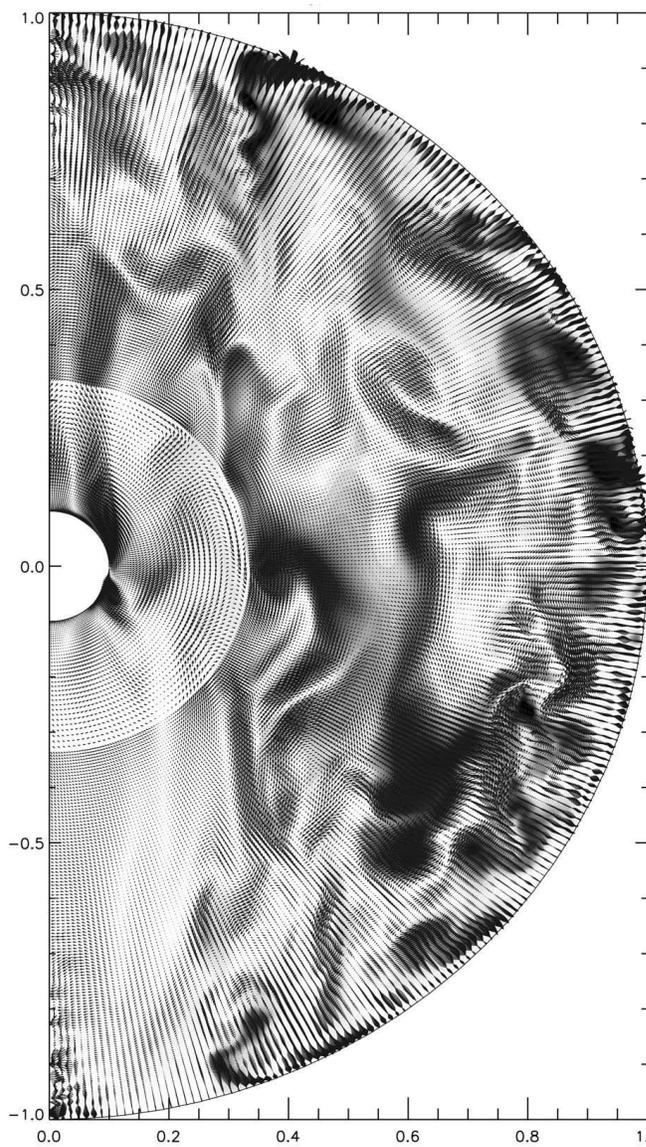} }
\vspace{-5mm}
\caption[]{Velocity field in a vertical cross-section through the 
pre-main-sequence model of convective processing of an initial 
large-scale field. Fewer grid points were plotted near the inner 
boundary for clarity. The time of this snap-shot is $t=0.001$ diffusion
times; the bottom of the convection zone is at $r=0.56$ at that time.}
\label{hdhordt}
\end{center}
\end{figure}

\section{Early convection\label{convection}}
The question of the origin of the magnetic energy contained in an Ap star 
and the selection problem discriminating normal A stars from magnetic ones 
certainly requires models and simulations of the pre-main-sequence evolution 
of intermediate-mass stars. We expect fully convective phases for stars up 
to about 2.5 M$_\odot$ (Alecian, this issue).

A mean-field dynamo model in a convection zone  whose depth diminishes 
with time was set up by Kitchatinov et al. (2001). Their last section 
addresses the remaining  rms field in the radiative zone below the 
convection zone. Because of the non-axisymmetry of the dynamo solutions 
of deep convection zones, the fields remaining in the radiative zone are 
of the order of a gauss  only.

Here we tried to follow the processing of an initial magnetic field in a 
direct numerical simulation including convection. Two basic initial 
conditions are compared: a homogeneous magnetic field $B_0$ parallel to 
the rotation axis, and a homogeneous magnetic field perpendicular to the 
rotation axis, which we will call ``equatorial'' in the following. In 
other words, if we imagine a Cartesian coordinate system $(x,y,z)$ in 
which the $z$-axis is aligned with the rotation axis of the star, one 
of the initial fields is $(0,0,B_0)$, and the other is $(B_0,0,0)$.

An almost full spherical computational domain is used (inner radius 0.1 
stellar radii) while the MHD equations are reduced to the Boussinesq 
approximation to be able to at least tentatively cover evolutionary 
time-scales with the simulations. The dimensionless equations solved are 
\begin{eqnarray*}
\frac{\partial \vec{u}}{\partial t} &=& - (\vec{u}\cdot \nabla) \vec{u} - \nabla  p +(\nabla\times \vec B)\times\vec B + {\rm Ra}\zeta(r,t) \Theta \vec{r} + {\rm Pm} \Delta \vec{u}\nonumber\\
\frac{\partial \vec{B}}{\partial t} &=& \nabla \times (\vec{u} \times \vec{B}) + \Delta \vec{B}\nonumber\\
\frac{\partial \Theta}{\partial t} &=& - \vec{u}\cdot \nabla\Theta - \vec{u}\cdot \nabla T + \frac{{\rm Pm}}{{\rm Pr}} \Delta \Theta ,
\label{1}
\end{eqnarray*}
where $\vec{u}$, $\vec{B}$, $\Theta$, and $p$ are the velocity, magnetic 
field, temperature fluctuations, and pressure, respectively. The equations
are solved with the MHD code by Hollerbach (2000) and 
are normalized by the radius of the star, $R_*$, and the magnetic diffusivity, 
$\eta$, leading to times measured in diffusion times, $\tau_{\rm diff}=R_*/\eta$ 
and the dimensionless numbers ${\rm Pm}=\nu/\eta$, ${\rm Pr}=\nu/\chi$, and 
${\rm Rm}=R_*^2 \Omega_*/\eta$, where $\nu$ is the viscosity, $\chi$ the 
thermal diffusivity and $\Omega_*$ the angular velocity of the star (at the pole). 
${\rm Ra}=g \alpha \Delta T d^3 /\chi \eta$ is a modified Rayleigh number 
with $g$ being the gravitational acceleration, $\alpha$ the thermal 
expansion coefficient, $\Delta T$ the temperature difference between the inner 
radius $r_{\rm i}$ and the outer radius $r_{\rm o}$, and $d$ is the gap width, 
$r_{\rm o}-r_{\rm i}$. The temperature fluctuations measure the deviations from 
the purely conductive profile $T=(1-r_{\rm o}/r)/(1-r_{\rm o}/r_{\rm i})$. The
linear terms are solved implicitly in a spectral space, while nonlinear terms
computed in real space on a grid with 240, 240, and 102 points in the radial,
latitudinal, and azimuthal directions, respectively.

The buoyancy is a function of radius and time 
being $\zeta(r,t) = 1/2 + \{1+{\rm erf}[(r-r_{\rm c})/0.02]\}$
where $r_{\rm c}(t)$ is the radius of the bottom 
of the convection zone. It depends on time by $r_{\rm c}(t)=(t/t_0)^{1/4}$ 
where $t_0=0.01 \tau_{\rm diff}$, which is the time by which the outer 
convection zone disappeared. The Rayleigh number in the convection zone was
${\rm Ra} = 2\cdot 10^8$. We do not simulate the onset of the core convection.

The magnetic Reynolds number ${\rm Rm}$ enters the equations by setting an 
initial rotation to the velocity field for which ${\rm Rm}=5\cdot 10^4$, while
the exact dimensionless formulation is a differential rotation of the form 
$u_\phi(t=0) = {\rm Rm}\,r\sin\theta/\sqrt{1+(r\sin\theta)^4}$, i.e. constant rotation on 
cylinders. Note that realistic stellar ${\rm Rm}$ and ${\rm Ra}$ are orders of 
magnitude higher, meaning that global numerical simulations always assume 
much larger diffusivities than the microscopic plasma diffusivities in 
stars. This is typically interpreted as an unresolved turbulence below 
the resolution scale of the simulations. However, rotating, stratified 
turbulence has not only diffusive effects on the large-scale quantities.
This is why such an argumentation is very weak, and we may miss substantial
physics when not resolving the smallest scales. Computing facilities are
very far away from being able to resolve these scale, unfortunately.
With this warning being kept in mind, let us see what is coming out of the
simulations anyway. 

\begin{figure}
\begin{center}
\hbox{ \includegraphics[width=9.8cm]{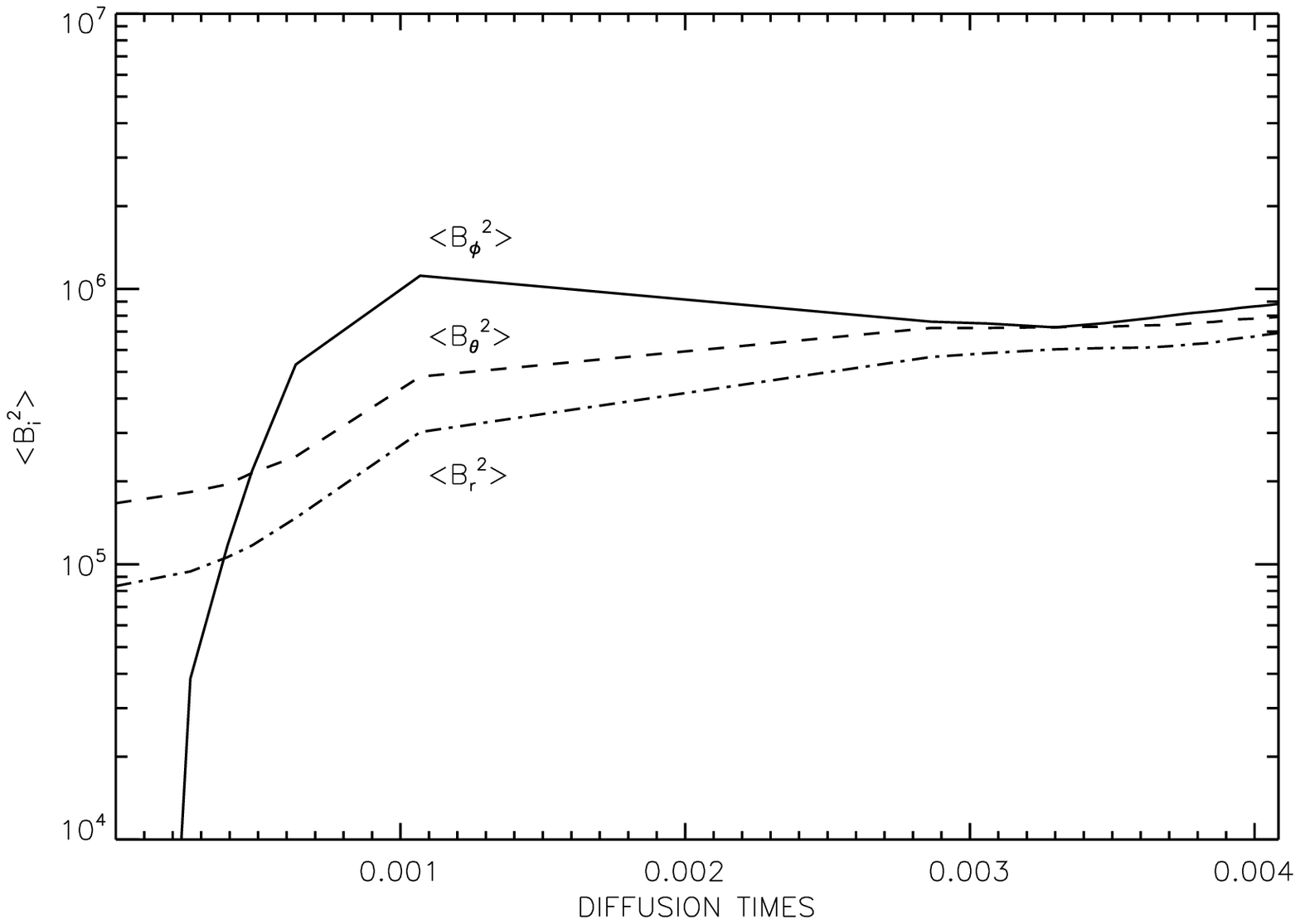} }
\hbox{ \includegraphics[width=9.8cm]{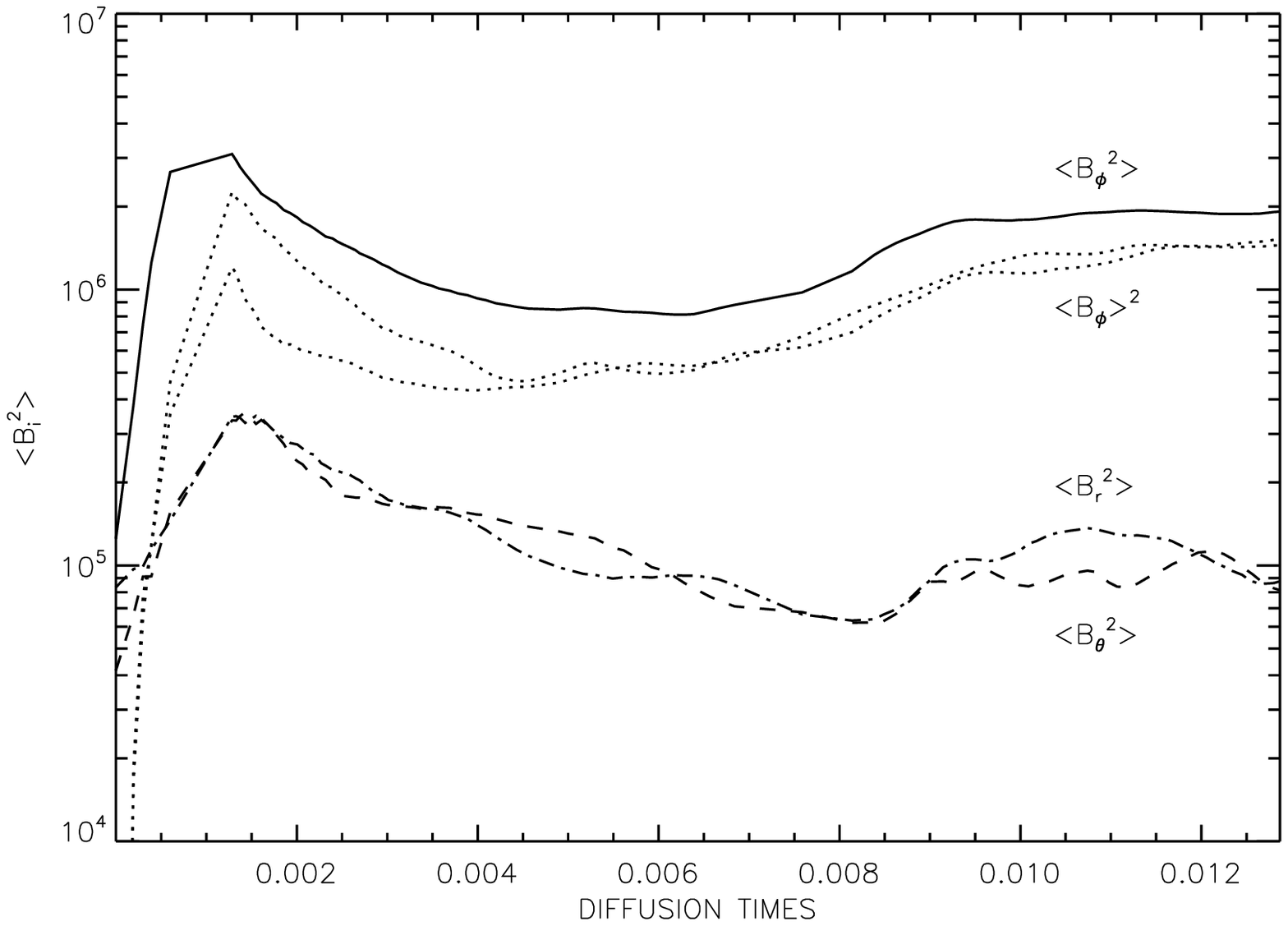} }
\vspace{-5mm}
\caption[]{Average unsigned magnetic field in the interior of the 
computational domain versus time for an initially vertical homogeneous 
magnetic field (top) and an initially horizontal equatorial magnetic 
field (bottom). The lower graph also shows the average signed field 
for the northern and southern hemispheres as dotted lines.}
\label{components}
\end{center}
\end{figure}

Figure~\ref{components} shows the temporal evolution of the magnetic field 
components in the computational domain averaged aver the volume at $0.1<r<0.5$ 
for a homogeneous vertical initial field and a homogeneous equatorial initial 
field. The bottom of the convection zone crosses the upper radial limit at $t=0.000625$. 
Note the different time-scale of the two graphs since it was much more difficult 
to run the simulation for the vertical initial field which in fact has not yet 
reached the time when the convection zone vanishes entirely ($t=0.01$). The 
average kinetic energy density is $5\cdot 10^8$, for comparison, due to the 
initial rotation with the total angular momentum being conserved by the 
numerical scheme.

The lower graph of Fig.~\ref{components} also shows the squares of the {\em signed\/} 
averages of $B_\phi$ in the two hemispheres indicating a substantial large-scale
field. The pure winding-up of the equatorial field would lead to zero averages
of the signed $B_\phi$ because of the non-axisymmetry (and it is in fact zero 
in the beginning while the rms is not).

The Alfv\'en speed of the internal magnetic field reaches roughly 7\% of the
rotational velocity which is, for a star rotating with 20~km/s at 0.5 stellar
radii (as a rough guess from the five magnetic HAeBe stars from Alecian et 
al.\ 2013b), about 200~kG. Note, however, that the initial field was already
50~kG in this scaling. Still, the storage of a factor four stronger field in
the radiative zone is a promising result given the very crude character of
the simulations presented here. The 200~kG fields may already be strong 
enough to explain the kGauss fields emerging on the surfaces of Ap stars
as observed.

The interesting outcome here is that the vertical initial field leads to
equal energies in the three magnetic field components, while the equatorial
initial field leads to a dominance of the azimuthal field. Combinations
of equally strong poloidal and toroidal fields are known to have the highest
stability, whereas purely toroidal fields will become unstable more easily.
There is no unique link between our energy plot and poloidal and toroidal 
fields, but at least it is an indication that the initial magnetic field
orientation gives a discriminating criterion for the later evolution into
stars with magnetic surfaces and stars with very weak or zero fields.

\section{Outlook}

We would like to stress the importance of studying the pre-main-sequence 
evolution of intermediate-mass stars in terms of convective processing of
magnetic fields and possible remnants in the emerging radiative zones.
I am sure we will find dynamo fields in the youngest Herbig stars which
need to be linked to the presence and absence of magnetic fields on main-squence
stars.

It will be interesting to improve the observations of the differential
rotation of A stars as compared to Ap stars at different evolutionary
stages. Ap stars appear to have no surface differential rotation 
(e.g. L\"uftinger et al. 2010). Normal A stars showed no surface
differential rotation in the Fourier analysis of spectral line profiles
(Ammler-von Eiff \& Reiners 2012) beyond 7400~K, while Kepler light 
curves indicate differential rotation of up to 15 per cent (Balona 2013).
Since the presence of internal large-scale magnetic fields has a great
impact on the differential rotation of a radiative star, the precise
knowledge of the rotation profiles, especially for young A stars, would 
be highly beneficial.

\bigskip
{\it Acknowledgements.} The author would like to thank the organizers of 
the conference for the kind invitation.


\begin{thebibliography}{99}

\bibitem{alec2013a}   
{Alecian E.,  Neiner C.,  Mathis S., et al.} 2013a, A\&A, 549, L8

\bibitem{alec2013b}
{Alecian E., Wade G.A., Catala, C., et al.} 2013b, MNRAS, 429, 1001  

\bibitem{alec2013c}
{Alecian E., Wade G. A., Catala C., et al.} 2013c, MNRAS, 429, 1027  

\bibitem{ammler}
{Ammler-von Eiff M., Reiners A.} 2012, A\&A, 542, A116 

\bibitem{arlt08}
{Arlt R.} 2008, Contributions of the Astronomical Observatory Skalnat\'{e} Pleso, 38, 163

\bibitem{arlt+rue11a}
{Arlt R., R\"udiger G.} 2011a, MNRAS, 412, 107

\bibitem{arlt+rue11b}
{Arlt R., R\"udiger G.} 2011b, AN, 332, 70

\bibitem{balona}
{Balona, L.} 2013, MNRAS, 431, 2240

\bibitem{b+n06}
{Braithwaite J., Nordlund \AA.} 2006, A\&A, 450, 1077

\bibitem{b07}
{Braithwaite J.} 2008, MNRAS, 386, 1947

\bibitem{b09}
{Braithwaite J.} 2009, MNRAS, 397, 763

\bibitem{brun05}
{Brun A. S., Browning M. K., Toomre J.} 2005, ApJ, 629, 461

\bibitem{ch+kend}
{Chandrasekhar S., Kendall P. C.} 1957, ApJ, 126, 457

\bibitem{dud+j89}
{Dudley M. L., James R. W.} 1989, Royal Society (London), Proceedings, Series A, 425, 407 

\bibitem{fe09}
{Featherstone N. A., Browning M. K., Brun A. S., Toomre J.} 2009, ApJ, 705, 1000

\bibitem{holl}
{Hollerbach R.} 2000, Int. J. Num. Meth. Fluids, 32, 773

\bibitem{kit01}
{Kitchatinov, L. L., Jardine, M., Collier Cameron, A.} 2001, A\&A, 374, 250

\bibitem{luftinger}
{L\"uftinger T., Fr\"ohlich H.-E., Weiss W. W., et al.} A\&A, 509, A43

\bibitem{Sp62}
{Spitzer L.} 1962, Physics of fully ionized plasmas. Interscience Publ., New York, London

\bibitem{st2000}
{St\c{e}pie\'{n} K.} 2000, A\&A, 353, 227

\bibitem{sz+arlt13}
{Szklarski J., Arlt R.} 2013, A\&A, 550, A94

\bibitem{wright}
{Wright G. A. E.} 1973, MNRAS, 162, 339

\end{thebibliography}
\end{document}